\documentclass{vldb}

\usepackage{microtype}
\usepackage{graphicx}
\usepackage[tight]{subfigure}

\usepackage{paralist}
\usepackage{pifont}
\usepackage{multirow}
\usepackage{amsmath}
\usepackage{amssymb}

\usepackage[linesnumbered,ruled,vlined,boxed]{algorithm2e}
\usepackage{fancyvrb}
\usepackage{listings}
\usepackage{url}
\usepackage{balance}

\usepackage{hyperref}
\hypersetup{
     colorlinks   = true,
     citecolor    = blue,
     linkcolor    = red,
     urlcolor     = blue
}

\graphicspath{{./pics/}}

\def\invpiclable{\vspace*{-0.25cm}}

\def\inv{\vspace*{-0.18cm}}

\newcommand{\commentout}[1]{}

\begin{document}

\title{\center Effective Spatial Data Partitioning for Scalable Query Processing}
\numberofauthors{3} 

\author{
%
%
\alignauthor
Ablimit Aji\\
       \affaddr{HP Labs}\\
       \affaddr{Palo Alto, California, USA}\\
       \email{ablimit@hp.com}
\alignauthor
Hoang Vo\\
       \affaddr{Emory University}\\
       \affaddr{Atlanta, Georgia, USA}\\
       \email{hoang.vo@emory.edu}
\alignauthor
Fusheng Wang\\
       \affaddr{Stony Brook University}\\
       \affaddr{Stony Brook,New York, USA}\\
       \email{fusheng.wang@stonybrook.edu}
}

\maketitle


\begin{abstract}
Recently, MapReduce based spatial query systems have emerged as a 
cost effective and scalable solution to large scale spatial data processing and analytics. 
MapReduce based systems achieve massive scalability by partitioning the data 
and running query tasks on those partitions in parallel. 
Therefore, effective data partitioning is critical for task 
parallelization, load balancing, and directly affects system performance. 
However, several pitfalls of spatial data partitioning make this task particularly challenging. 
First, data skew is very common in spatial applications. 
To achieve best query performance, data skew need to be reduced to the minimum.  
Second, spatial partitioning approaches generate boundary objects that cross 
multiple partitions, and add extra query processing overhead. 
Therefore, boundary objects need to be minimized.  
Third, the high computational complexity of spatial partitioning algorithms combined 
with massive amounts of data require an efficient approach for partitioning to achieve overall fast query response.
In this paper, we provide a systematic evaluation of multiple spatial
partitioning methods with a set of different partitioning strategies, and study their implications on
the performance of MapReduce based spatial queries.
We also study sampling based partitioning methods and their impact on queries,
and propose several MapReduce based high performance spatial partitioning methods.
The main objective of our work is to provide a comprehensive guidance for optimal spatial data partitioning to
support scalable and fast spatial data processing in distributed computing environments such as MapReduce.
The algorithms developed in this work are open source and can be easily
integrated into different high performance spatial data processing systems.
\end{abstract}

\section{Introduction}\label{sec:introduction}
The proliferation of ubiquitous positioning technology, mobile devices, and
the rapid improvement of high resolution data acquisition technologies enabled
us to collect massive amounts of \emph{spatial data} in a way that was never
before possible. The volume and velocity of data only increase significantly
as we shift towards the Internet of Things paradigm in which devices have spatial awareness
and produce data while interacting with each other. As science and businesses are becoming increasingly data-driven,
timely analysis and management of such data is of utmost importance to 
data owners. A wide spectrum of applications and scientific disciplines
such as GIS, Location Based Social Networks (LBSN), neuroscience \cite{ailamaki10sci}, medical imaging
\cite{wang11pais} and astronomy \cite{crossmatchingsky},
can benefit from an efficient spatial query system to cope with the challenges of
Spatial Big Data.

To effectively store, manage and process such large amounts of spatial data, a
scalable distributed data management system is essential.
Recently, the MapReduce framework \cite{dean2008mapreduce} has become the de
facto standard for handling large scale data processing tasks, and it has
many salient features such as massive scalability, fault-tolerance, easy programmability and low deployment cost.
With the success of MapReduce, a number of spatial query systems
\cite{aji12imaginggis,lu2013parallel,ray2013parallel} and
frameworks \cite{aji13hadoopgis,eldawy2013demonstration} have emerged to enable large scale spatial
query processing on MapReduce and cloud platforms.


Data partitioning is a powerful mechanism for improving efficiency of
data management systems, and it is a standard feature in modern database
systems. In fact, state-of-the-art systems employ a shared-nothing
architecture~\cite{stonebraker1986case}, and both MapReduce and parallel DBMS are examples of such architecture.
Aside from the fact that data partitioning improves the overall manageability of large datasets,
it improves query performance in two ways.
First, partitioning the data into smaller units enables processing of a query in parallel,
and henceforth the improved throughput. Second, with a proper partitioning schema,
I/O can be significantly reduced by only scanning a few partitions that contain relevant data
to answer the query.
Therefore, a partitioning approach -- that evenly distributes the data across nodes
and facilitates parallel processing -- is essential for achieving fast query response
and optimal system performance.

Spatial data partitioning, however, is particularly challenging due to
several pitfalls that are endemic to spatial data and query processing.

\noindent {\bf Spatial Data Skew}. Data skew is very common and severe in
spatial applications. For example, in microscopic pathology imaging scenario, tumorous
tissues contain far more spatial objects (segmented cells), whereas cells are more
evenly distributed in healthy tissues. In geospatial applications (e.g.,
OpenStreetMap) some countries and regions have more detailed mapping
information due to the enthusiastic data contributors. 
For example,  if OpenStreetMap is partitioned into 1000 x 1000 fixed size tiles, the number of objects contained in the most skewed tile is nearly three orders of magnitude more than the one in an average tile.
Needless to say, data skew is detrimental to the
query performance \cite{sowell13experimental}
and curtails system scalability \cite{patel97building}. Therefore, to
achieve the best query performance, a spatial partition approach should try to 
avoid a skewed partitioning whenever it is possible.


\noindent {\bf Boundary Objects}. Spatial partitioning approaches generate
boundary objects that cross multiple partitions, thus violating the partition
independence. As spatial objects have complex boundary and extent, imposing
a rectangular region based partitioning on sufficiently large dataset would
most certainly produce objects that cross multiple partition boundary.
Spatial query processing algorithms get around the boundary problem by using a
\emph{replicate-and-filter} approach \cite{patel97building,zhou1998data} in
which boundary objects are replicated to multiple spatial partitions, and side
effects of such replication is remedied by filtering the duplicates at the end
of the query processing phase. This process adds extra query processing
overhead which increases along with the volume of boundary objects. Therefore,
a good spatial partitioning approach should aim to minimize the number of
boundary objects.

\noindent{\bf Performance}. Spatial partitioning algorithms are 
expensive to compute compared to the conventional one dimensional table
partitioning algorithms, such as hash and range partitioning, that can be
done quickly on the fly.
The multidimensional nature of spatial data entails that most spatial operators
are of linear time complexity. The high computational complexity combined with
massive amounts of data require an efficient approach for spatial partitioning
to achieve overall fast query response. This is in particularly important for
spatial-temporal data where new spatial data has to be partitioned and processed
in a timely fashion.

To the best of our knowledge, no spatial database system provides a graceful
approach to spatial partitioning. Previously, Paradise \cite{patel97building}
-- a parallel spatial database system -- used a regular fixed grid
partitioning for parallel join processing. Fixed grid partitioning is the
basis of many spatial algorithms and it is easy to compute. However, as
mentioned in the original work, fixed grid approach suffers from both
\emph{data skew} problem and \emph{boundary object} problem.

In the relevant research literature, some of those challenges are given some
attention in various contexts. However, in most cases the problem is not
fully explored, or circumvented by providing domain specific ad-hoc fixes.
To fill those gaps, a systematic and detailed study of spatial partition
approaches for parallel spatial query processing is needed. In this paper,
we provide a detailed study of a set of six spatial partitioning approaches
within a MapReduce based spatial query processing framework
\cite{aji13hadoopgis}. The approaches described here can be reused
for any other distributed or parallel spatial query processing systems beyond MapReduce that
can take advantage of data partitioning. We provide parallelization strategies to improve the
partitioning algorithm efficiency which can improve the efficiency of a number
of tasks such as data loading and ad-hoc query processing.
In summary, our main contributions are as follows:
\begin{asparaenum}
  \item To the best of our knowledge, this is the first formal attempt to
    introduce and address the spatial data partitioning problem for parallel
    query processing.
  \item We present six spatial partitioning algorithms in details, and provide
  a general classification of those approaches along three dimensions.
\item We systematically study various properties of the presented spatial
  partitioning algorithms and their effects on query performance, and provide
  a comprehensive empirical evaluation on two large scale real-world datasets.
  \item We propose MapReduce based algorithms for parallel spatial
    partitioning, and evaluate their performance in details.
\end{asparaenum}


\section{Background} \label{sec:background}


\subsection{Spatial Query Processing with MapReduce}
Recently, several MapReduce based spatial query systems
\cite{aji12imaginggis,eldawy2013demonstration} have emerged to support
scalable spatial query processing on large datasets. While these systems may
vary in implementation details and at the query language layer, conceptually
they are very similar. Algorithm \ref{algo:framework} sketches out HadoopGIS
-- a general MapReduce based spatial query processing framework~\cite{aji13hadoopgis}.
As the algorithm shows, data is spatially partitioned and staged to HDFS;
spatial queries are expressed as a set of operators that can be translated to
MapReduce tasks during runtime. Tasks run on the partitioned input for
parallel query processing. Queries are implicitly
parallelized through MapReduce, and a \emph{tile} (as spatial partitioning closely
resembles tiling of two dimensional space, we use tile and spatial
partition interchangeably hereafter) is the parallelization unit
that a Mapper/Reducer can process independently.

\begin{algorithm}[bht]
\SetKw{KwTo}{in}
A. Data/space partitioning\;

B. Staging of partitioned data to HDFS\;

C. Pre-query processing (optional)\;

D. \For{$tile$ \KwTo $input$}{
     Index building for objects in the tile\;
     Tile based spatial querying processing\;
}

E. Boundary object handling\;

F. Post-query processing (optional)\;

\caption{MapReduce based spatial query processing framework}
\label{algo:framework}
\end{algorithm}

\begin{figure}[h]
 \centering
 \includegraphics[width=0.9\linewidth]{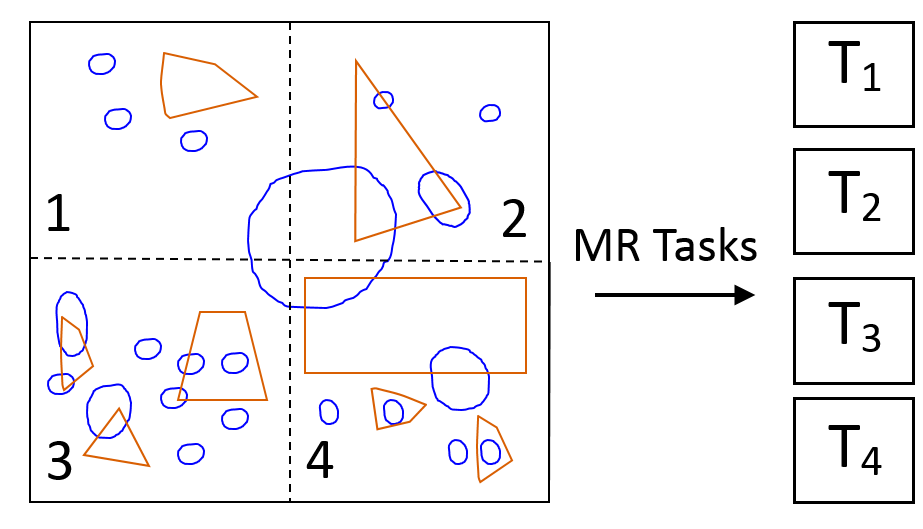}
\caption{An example of MapReduce based spatial query parallelization}
\label{fig:hadoopgisexample}
\end{figure}

Fig. \ref{fig:hadoopgisexample} shows a simple example where the dataset
is partitioned into four tiles (dotted lines depict partition boundary).
To process a spatial join query such as
\emph{find object pairs that intersect with each other from two datasets}, a single MapReduce
job can be started where each tile is processed by a single mapper
($T_{1},T_{2},T_{3},T_{4}$).
In this example, tile $3$ contains more objects than other tiles,
and consequently requires more processing time. As a result, the corresponding
MapReduce task ($T_{3}$) becomes a \emph{straggler task} -- a performance bottleneck in
MapReduce based queries.

Given a spatial dataset $R$, ideally, we would like to derive a spatial decomposition of
$R= \cup_{i=1}^{k}R_{i}$ where $R_{i} \cap R_{j} =\varnothing$ for $i \neq j$,
such that query tasks can be performed on each partition $R_{i}$ independently in parallel.

\subsection{Boundary Objects}
There is one specific problem that is endemic to spatial partitioning -- {\em boundary objects}.
Due to their multidimensional nature, some spatial objects may span more than one
partition. For example, in Fig. \ref{fig:hadoopgisexample} the big round
object in the middle crosses all tile boundaries.  As a result, $R_{i} \cap R_{j} \neq \varnothing$
for $i \neq j$, and it requires the spatial query processing framework to be able to handle such cases.

Parallel spatial query processing algorithms remedy the boundary object problem in two ways.
Namely {\em multi-assignment, single-join} (MASJ) and {\em single-assignment, multi-join}
(SAMJ) \cite{zhou1998data,lo1996spatialhash}. In MASJ approach, each boundary
object is replicated to each tile that overlaps with the object. During
the query processing phase, each partition is processed only once without
considering the boundary objects. Then a {\em de-duplication} step is
initiated to remove the redundancies that resulted from the replication.
However, in SAMJ approach, each boundary object is only assigned to one tile.
Therefore, during the query processing phase, each tile is processed
multiple times to account for the boundary objects.

Both approaches introduce extra query processing overhead. In MASJ, the replication of boundary objects incurs
extra storage cost and computation cost. In SAMJ, however, only extra
computation cost is incurred by processing the same partition multiple
times. Hadoop-GIS takes the MASJ approach \cite{aji13hadoopgis} and the original
work pointed out that:
\begin{inparaenum}[(a)]
\item In practice, the MASJ approach is proven to be significantly efficient than
the SAMJ approach \cite{zhou1998data};
\item MASJ approach allows higher degree of parallelization
such that, for large datasets, the query processing efficiency can be greatly improved,
and de-duplication cost can be diminishingly small.
\end{inparaenum}

\subsection{Query Processing Cost Model}
In Hadoop-GIS framework, the cost of processing a query includes non-boundary
objects processing cost, duplicated boundary objects processing cost, and
object de-duplication cost. A modeling approach can help us better understand
overall query processing overhead and provide principled guidelines for
optimizing spatial partitioning for optimal query performance.

Given two datasets $R$ and $S$ (round objects and polygonal objects in Fig \ref{fig:hadoopgisexample}), a spatial join query finds all the pairs
$r_{i}\in R, s_{j}\in S$ that satisfies a spatial topology relationship
$F(r_{i},s_{j}) =1$. Selection of the spatial topology
function can be arbitrary and without loss of generality,
we use \verb"st_intersects" as an example throughout the paper. Let us assume that, datasets are merged
and co-partitioned with a partition schema which results in partitions of
$R= \cup_{i=1}^{k}R_{i}$ and $S= \cup_{i=1}^{k}S_{i}$. Following the MapReduce based query processing
framework , we have:
\begin{equation}\label{eq:pjoin}
R \stackrel{F}{\bowtie} S = \bigcup^{k}_{i=1} R_{i}\stackrel{F}{\bowtie} S_{i}
\end{equation}
A few assumptions can help us simplify the analysis.
\begin{inparaenum}[(a)]
\item Each dataset follows a uniform distribution.
Consequently, ignoring boundary objects, each partition contains roughly $|R|/k$ and $|S|/k$ objects.
\item For each partition, the fraction of boundary objects is $\alpha$. Hence, each
partition contains $(1+\alpha)*|R|/k$ objects as a result of boundary object replication.
\item Overall query processing cost is the sum of partitioned query processing cost $C_{1}$,
and data deduplication cost $C_{2}$ which is a fixed amount ($\beta(|R|+|S|$) that bound by the dataset size \cite{aji13hadoopgis}.
\end{inparaenum}
After we plug in the cost functions into the equation \eqref{eq:pjoin}, the
overall query processing cost is:
\begin{equation}\label{eq:cost}
\begin{split}
C(R \stackrel{F}{\bowtie} S) &= \sum^{k}_{i=1} C_{1}(R_{i}\stackrel{F}{\bowtie} S_{i}) + C_{2} \notag\\
& = \sum^{k}_{i=1}\frac{(1+\alpha)|R|}{k}\frac{(1+\alpha)|S|}{k} + \beta(|R|+|S|) \notag\\
& = \frac{(1+\alpha)^2|R||S|}{k} +\beta(|R|+|S|)
\end{split}
\end{equation}

The cost model provides us an important insight -- partition granularity is a double-edged sword.
On one hand, a finer level of partitioning (larger $k$) improves the query performance.
On the other hand, a finer level of partitioning generates a larger
fraction of boundary objects (larger $\alpha$), and consequently it is detrimental to the query performance.
Clearly, there is a sweet spot for the partition granularity which yields the best query performance.
Finding the optimal partition granularity is non-trivial as it depends on the
dataset characteristics and query type, and we plan to explore this problem separately in our future work.

\section{Classification of Spatial Partition Algorithms} \label{sec:class}
In this paper we study six spatial partition algorithms that are representative
of different classes of approaches.
Before we delve into the technical details, it would be more interesting to
give a high-level view to help understand how these algorithms are related,
and what their major differences are. Here, we attempt to categorize those
algorithms along three dimensions, and Table \ref{table:category} summarizes
such classification. The algorithmic details will be discussed in next section.
\begin{table*}[htb]
\centering
\begin{tabular}{|c|c||c|c|c|c|c|c|}
\hline
{\bf Dimension}                     & {\bf Category}  &  BSP &  FG  &    SLC  &  BOS & STR  &  HC \\
\hline
\hline
\multirow{2}{*}{Partition Boundary} & overlapping     &            &            &            &            &
\checkmark & \checkmark \\ \cline{2-8}
                                    & non-overlapping & \checkmark & \checkmark & \checkmark & \checkmark & & \\
\hline
\multirow{2}{*}{Search Strategy}    & top-down        & \checkmark &     NA       &            &            &
& \\ \cline{2-8}
                                    & bottom-up       &            &     NA       & \checkmark & \checkmark &
                                    \checkmark        & \checkmark \\
\hline

\multirow{2}{*}{Split Criterion} & space-oriented   &\checkmark  & \checkmark &            &            &
& \\ \cline{2-8}
                                    & data-oriented   &  &            & \checkmark & \checkmark &
                                    \checkmark        & \checkmark \\
\hline
\end{tabular}
\caption{A general classification of spatial partition algorithms. BSP: Binary split 
partitioning, FG: Fixed grid partitioning, SLC: Strip partitioning, BOS: Boundary
optimized strip partitioning, STR: sort-tile-recursive partitioning, HC: Hilbert curve partitioning.}
\label{table:category}
\end{table*}
\subsection{Partition Boundary}
We start with whether the spatial partition boundaries overlap with each other.

\noindent\textbf{Non-overlapping.} Algorithms in this category generate
spatial partitions of which boundaries do not overlap with each other.
Non-overlapping partitioning is ideal for most query processing tasks as it
does not incur any extra storage or computation overhead other than replicated
boundary objects. Due to the same reason, in this paper we mostly focus on
this class of algorithms which includes FG, BSP, SLC, and BOS.

\noindent\textbf{Overlapping.} Algorithms in this category relax the
non-overlapping boundary condition, and allow generated partitions to
overlap with each other. Most spatial index construction 
algorithms~\cite{samet2006foundations} are based 
on the similar idea, and the packing algorithms such as STR
\cite{leutenegger1997str} and Hilbert Curve \cite{kamel94hilbert} belong to
this class. Since the partitions may overlap with each other, some fraction
of objects would be present in multiple partitions. Those multi-partition
objects would be replicated and assigned to each of the overlapping
partitions. As a result, in this class of approaches the replication factor
$\alpha$ can be high which consequently increases the deduplication cost
factor $\beta$. However, if a \emph{good} partitioning can be \emph{quickly}
obtained by allowing the partitions to overlap, then the extra cost can be
compensated by the improved query performance.

\subsection{Search Strategy}
The second dimension we consider is the search strategy which focuses on how
the partitions are generated.

\noindent\textbf{Top-down.} This class of algorithms generate partitions in
top-down manner. Specifically, given a dataset and an expected partition
payload $b$ (number of objects assigned to that partition), a top-down
approach recursively splits the dataset into $k$ sub-partitions, and examines
if any sub-partitions has more than $b$ objects. If a sub-partition
has more than $b$ objects, then it will be further partitioned, until the payload
requirement is met. Most spatial indices are constructed using similar
procedure. While the value of the parameter $k$ can be chosen arbitrarily,
some specific values, such as $k=2$ (BSP) and $k=4$ (Quad-Tree), are used
more frequently in practice. Depending on the split criterion, this class of
algorithms can be implemented as either data-oriented or space-oriented, and
we describe these categories in the next subsection.

\noindent\textbf{Bottom-up.} Rather than generating partitions in a recursive
manner, this class of algorithms attempt to construct the final partitions as
early as possible. Such approach bears some resemblance to the spatial
packing algorithms. The general idea is to use proximity information of
spatial objects to group them into partitions. Since there is no spatial
proximity preserving total ordering for multi-dimensional objects, Space
Filling Curves are used to generate approximate one dimensional ordering.
Then, objects are packed into partitions by grouping them based on such
ordering.

\subsection{Partition Criterion}
Finally, splitting an oversized partition into smaller ones is a core
subroutine in spatial partitioning, and algorithms may have different
criterion for this task. For example, consider a simple case where a
partition with payload $w$ need to be partitioned into two sub-partitions.
There will be two strategies: space oriented, and data oriented.

\noindent\textbf{Space Oriented.} This class of algorithms generate
sub-partitions by \emph{spatially decomposing} the current partition boundary
into two equal sub-spaces. As the split decision is made solely based on the
space, this approach suffers from data skew. If the data distribution
is uniform, we would expect to get two sub-partitions where each of them has
a payload of roughly
$\displaystyle \frac{w}{2}$. However, if the data distribution is skewed, it
is possible that one of the subpartitions still contains large fraction of objects in
the original partitions, while the other contains only few objects.

\noindent\textbf{Data Oriented.} This class of algorithms generate
sub-partitions by finding a {\em cut} such that each resulting
sub-partitions contains roughly equal amounts of data ($\displaystyle \frac{w}{2}$).
The cut position is derived based on the distribution of data objects rather than splitting the space.
However, finding an optimal cut which generates an even partitioning requires significant
computational effort. Furthermore, the algorithms also need to be judicious
about the split position so that the number of boundary objects induced by
such split is not very large.

\section{Spatial Partition Algorithms}\label{sec:partition}
\subsection{Preliminaries}
We study the following partition problem: given a set of $d$-dimensional spatial objects
$R=\{r_{1},r_{2},.. r_{i} .. r_{n}\}$ $(|R|=N)$, a partition algorithm partitions $R$ into $k$ partitions
$P=\{p_{1},p_{2},.. p_{j} .. p_{k}\}$, where each partition is size bounded $|p_{j}| \leq b$,
and the number of partitions $k$ is minimized. Without loss of generality, we consider the case where $d=2$ and
a spatial object is approximated by its MBR (Minimum Bounding Rectangle),
and each rectangle is represented by $r_{i}=(x_{i},y_{i},u_{i},w_{i})$.

Partitioning of one dimensional data ($d =1$) has been extensively studied in
the past, and it is shown that the optimal solution can be obtained in
polynomial time \cite{jagadish1998optimal}. However, for higher dimensions,
even for a simple case $d=2$, the problem becomes intractable. Previously, a
simpler version of the problem, known as rectangle tiling, was studied. The
main objective of rectangle tiling is to partition a matrix of
integers into
tiles, and it was proven to be NP-Hard
\cite{fowler1981optimal,khanna1998approximating} for cases $d\geq 2$.

\begin{figure*}[tbh]
  \centering
  \subfigure[BSP]{\includegraphics[width=0.19\textwidth]{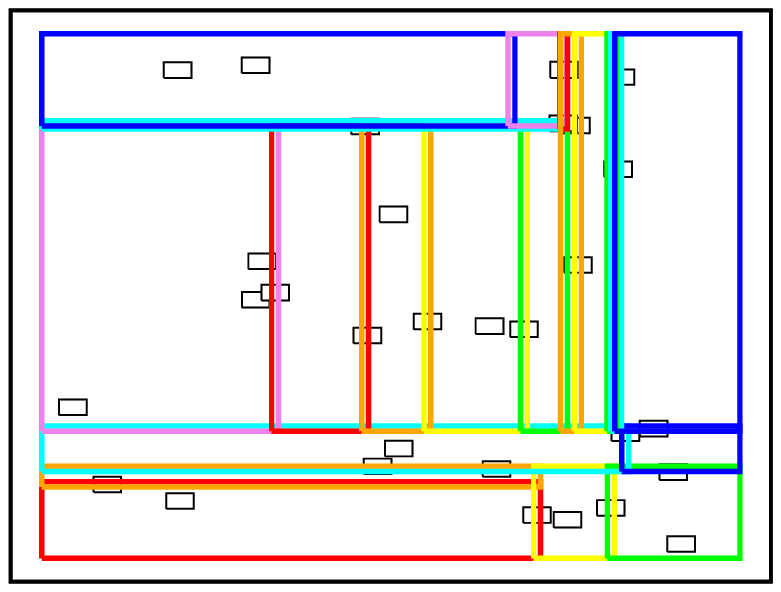}\label{fig:exbsp}}
  \subfigure[HC]{\includegraphics[width=0.19\textwidth]{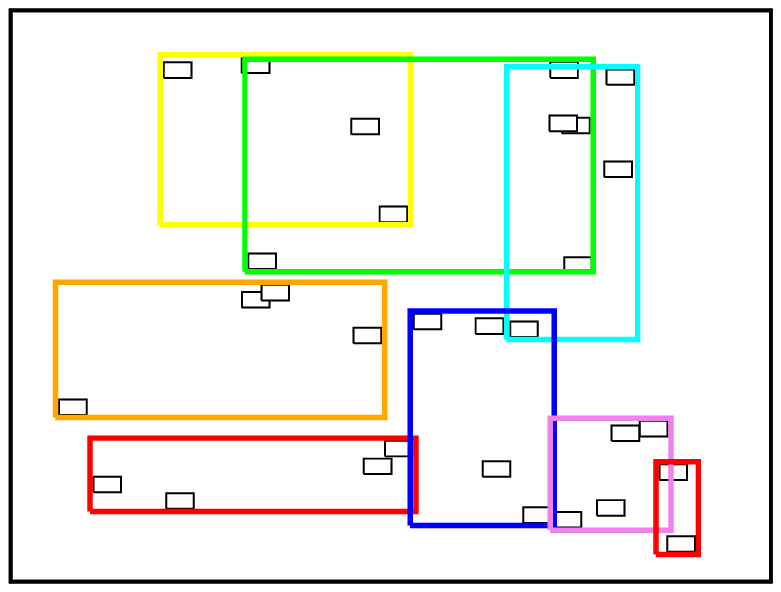}\label{fig:exhc}}
  \subfigure[SLC]{\includegraphics[width=0.19\textwidth]{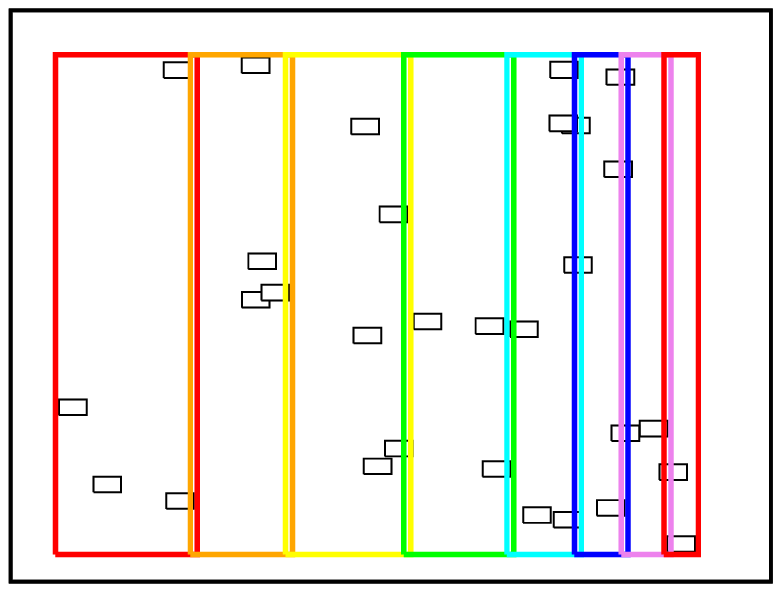}\label{fig:exst}}
  \subfigure[BOS]{\includegraphics[width=0.19\textwidth]{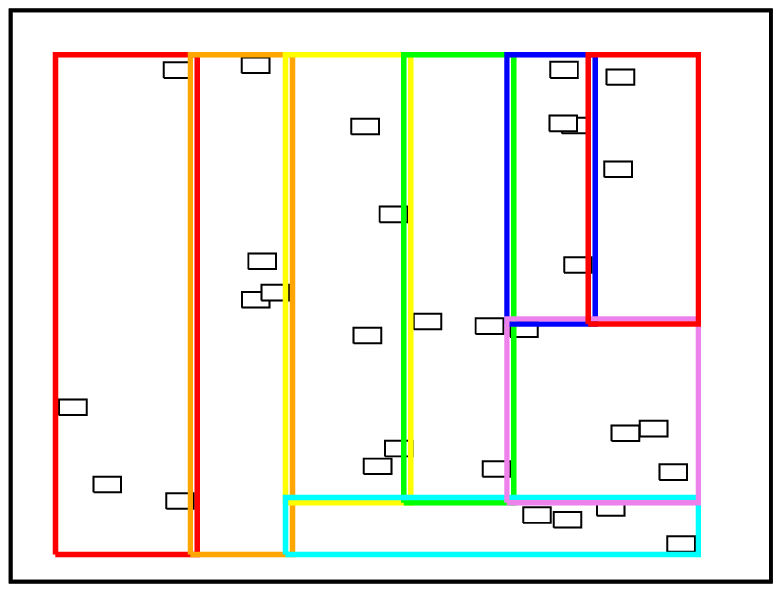}\label{fig:exrp}}
  \subfigure[STR]{\includegraphics[width=0.19\textwidth]{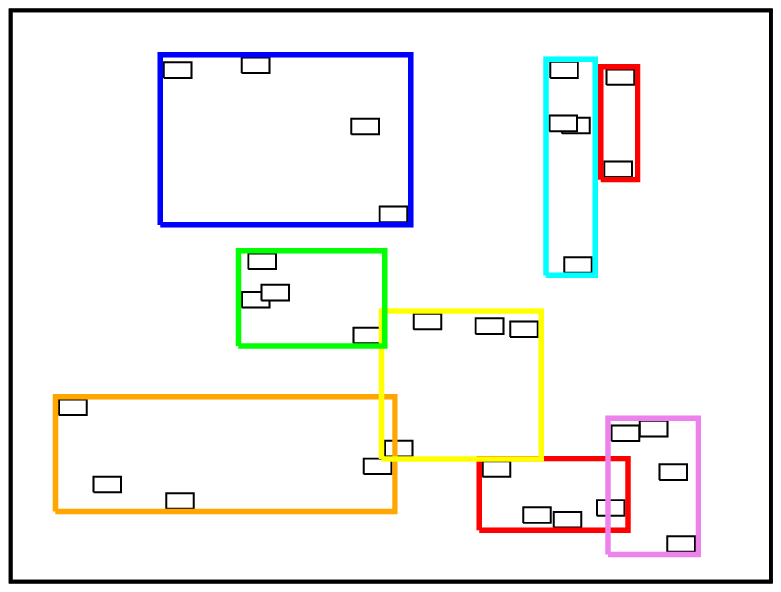}\label{fig:exstr}}
  \caption{Spatial partitions generated by different algorithms. 
  The bigger rectangles in colors represent partition boundaries, 
  and the small rectangles represent the spatial objects.}\invpiclable
  \label{fig:partitionexample}
\end{figure*}

\subsection{Methods and Details}

\noindent \textbf{Fixed Grid Partitioning (FG).}
Fixed grid partitioning is a simple space-oriented, non-overlapping
partitioning approach in which the spatial universe is partitioned into $k$
equal sized grids. A major assumption behind this approach is that data
follows a uniform distribution. Therefore, if the data distribution is close
to a uniform distribution, FixedGrid is expected to generate a reasonably good
partitioning. While the partition process is very simple, for the sake of clarity,
details of this approach are described in Algorithm \ref{algo:fixedgrid}.

\SetKwProg{Fn}{function}{\string:}{}
\SetKwFunction{In}{in}
\SetKw{KwTo}{in}

\begin{algorithm}[hbt]
  \SetAlgoLined
\KwIn{a set of spatial objects $R$ }
\KwIn{partition payload $b$ }
$m$ = $\lceil\sqrt{|R|/b}\phantom{.}\rceil$\;
$U$ = spatialUniverse($R$)\;
$G$ = split $U$ into $m$ by $m$ grid\;
\For{$r_{i}$  \KwTo $R$ $   $}
{
$g$  = grids intersects with $r_{i}$\;
assign $r_{i}$ to each grid in $g$\;
}
\caption{Fixed grid partition (FG) }
\label{algo:fixedgrid}
\end{algorithm}

\noindent \textbf{Binary Split Partitioning (BSP).}
Binary split partitioning is a top-down approach that generates partitions by
recursively dividing a given spatial partition into two non-overlapping
subpartitions until the payload requirement is met. Given a expected partition
payload $b$, BSP recursively creates subpartitions if the number of objects
inside a partition exceeds the specified payload (Algorithm \ref{algo:binarysplit}). The split point is chosen to
be the median of object centroids in that partition. The direction of the
split (horizontal or vertical) is dependent on the relative ratio of areas of
subpartitions. The split direction is chosen so that the relative area difference between children nodes are minimized based on a probabilistic expectation. 

\begin{algorithm}[hbt]
  \SetAlgoLined
\SetKw{KwTo}{to}
  \KwIn{a set of spatial objects $R$ }
  \KwIn{partition payload $b$ }

  $U$ = spatialUniverse($R$)\;
  \While {$r$ in  $R$}{
    $n$ = node($U$)\;
    addObject($n$, $r$)\;
  }
  \Fn{addObject{($n$,$r$)}}{
    \If{ $n$ is leafNode }{
	$n$.objectList.add($r$)\;
    }
	\If {size($n$.objectList) $\le c$} {
	   compute $median\_x$ and $median\_y$ split \;
	   $split$ = argmax(Product of children areas)\;
	
	   $child1$, $child2$ = children($n$, $split$)\;
	   \If {$child1$ intersects with $r$} {
	      addObject($child1$, $r$)\;
	   }
	   \If {$child2$ intersects with $r$} {
	      addObject($child2$, $r$)\;
	   }
	}
  }
  \caption{Binary split partition (BSP) }
  \label{algo:binarysplit}
\end{algorithm}

\noindent \textbf{Strip Partitioning (SLC).}
Strip partitioning is a non-overlapping, data oriented partitioning approach
that has some resemblance to \emph{slicing} a cake. In this approach, rather
than defining a fixed space, we slice off a rectangular region from the spatial
universe where each region contains approximately $b$ objects. Then similar
process is continued on the rest of the data and repeated until we generate
all the partitions. 
Details of this approach are described in Algorithm \ref{algo:strip}.
\inv 
\begin{algorithm}[bht]
  \SetAlgoLined
\SetKw{KwTo}{in}
\SetKwFunction{In}{in}
  \KwIn{a set of spatial objects $R$ }
  \KwIn{partition payload $b$ }
  \KwIn{partition dimension $d$ }

  \tcc{sort objects by mbr center in dimension $d$}
  sort ($R$,$d$)\;
  $U$ = spatialUniverse($R$)\;
  \While {$R$ is not empty $    $}{
    $s$ = cutStrip($U$, $R$, $b$)\;
    \For{$r_{i}$  \KwTo $R$ $   $}
    {
      \If{not $r_{i}$ intersects with $s$}
      {
       break\; 
      }
      assign $r_{i}$ to partition $s$ \;
        \If{$s$ contains $r_{i}$ }{
          remove $r_{i}$ from $R$\;
        }
    }
  }
  \caption{Strip partition (SLC) }
  \label{algo:strip}
   \inv \inv
\end{algorithm}

\noindent \textbf{Boundary Optimized Strip Partitioning (BOS).}
Algorithms described above do not explicitly consider the boundary object
problem, although the partition payload is guarenteed to be balanced. As a
result, we may still get a partitioning that are balanced but has a higher
deduplication cost. BOS is a boundary object aware extension of SLC that
minimizes the number of boundary objects while still generating a balanced
partitioning. While performing the strip based partitioning, BOS has two
dimensions ($d$ dimensions in general) to choose at each step. BOS calculates
the partitioning in both dimensions, and selects the one which induces
smaller number of boundary objects. Algorithm \ref{algo:bos} describes the
details of this approach.
\begin{algorithm}
  \SetAlgoLined
  \KwIn{a set of spatial objects $R$ }
  \KwIn{partition payload $b$ }

  $U$ = spatialUniverse($R$)\;
  \While {$R$ is not empty $    $}{
    \tcp{cost in dimension $x$}
    $cx$ = getCost($U$, $R$, $b$, $x$)\;
    \tcp{cost in dimension $y$}
    $cy$ = getCost($U$, $R$, $b$, $y$)\;

    \If{$cx \le cy$ }{
      strip partition in $x$ dimension\;
    }
    \Else{
      strip partition in $y$ dimension\;
    }
  }
  \caption{Boundary optimized strip partition (BOS) }
  \label{algo:bos}
   \inv \inv
\end{algorithm}

\noindent \textbf{Hilbert Curve Partitioning (HC).}
Space filling curves used in many application to obtain a locality preserving approximate total
ordering for multidimensional data. Commonly used space filling curves include
Z-curve, Gary-coded curve, and Hilbert curve. Among those approaches, Hilbert curve
is shown \cite{moon2001analysis} to have better clustering property for two dimensional
objects. In our implementation, we use Hilbert curve to map the centroid of
the spatial objects to obtain the curve value, and sort the dataset based on
the curve value. Then, we group each consecutive $b$ objects together to form
a spatial partition, and the union of their extent is the final partition
boundary.

\noindent \textbf{Sort-Tile-Recursive (STR) Partition.}
Packing spatial objects for bulk loading spatial index can be regarded
as a ``mini-partition'' step. Most often the leaf nodes are pre-packed in
order to generate low level nodes of the index, and higher level index nodes
are constructed from the leaf nodes. Similarly, we can use packing algorithms to
generated spatial partitions such that we only generate the lowest level index
nodes, and the node boundary serves as partition boundary. STR
\cite{leutenegger1997str} first partitions the spatial universe into large
vertical strips, then each strip is further partitioned in the horizontal
direction. Algorithm \ref{algo:str} illustrates the partition process.
\begin{algorithm}
  \SetAlgoLined
  \SetKw{KwTo}{to}
  \KwIn{a set of spatial objects $R$ }
  \KwIn{partition payload $b$ }

  $m$ = $\lceil\sqrt{|R|/b}\phantom{.}\rceil$\;

  \tcp{$m$ strips in dimension $x$}
  $S$ = stripPartition ($R$, $x$)\;

  \For{$i\leftarrow 1$ \KwTo $m$}{
    \tcp{$m$ strips in dimension $y$}
    $t$ = stripPartition ($S[i]$, $y$)\;
  }
  \caption{Sort-Tile-Recursive partition (STR) }
  \label{algo:str}
   \inv \inv
\end{algorithm}

Figure \ref{fig:partitionexample} shows a simple example in which 
a set of $32$ randomly distributed spatial objects are 
partitioned with different spatial partition algorithms we described above.

\section{Spatial Partition Efficiency}\label{sec:partitionefficiency}
Optimal spatial partitioning is NP-hard. For spatial query
processing tasks, performance of a query on an optimal partition layout may
not be so different than the one on a suboptimal partitioning.  Therefore,
finding a reasonably well partitioning in an efficient manner has
practical implications for many real world applications. In a spatial data
warehousing scenario, the underlying dataset is large and relatively stable,
and queries run on the same dataset many times. In such case, an approach that
produces a balanced partitioning but requires significant computational
resources may be acceptable as it improves the query performance in the long
run. However, in some other application scenarios such as scientific data
exploration and simulation, queries consume large amounts of intermediate data
that are generated quickly, and most queries run only once as the data being
generated. In such cases, a fast partitioning algorithm is critical for
achieving overall fast query response. In this paper we explore two different
approaches towards improving spatial data partitioning efficiency, namely
\emph{parallel} spatial partitioning and partitioning with \emph{sampling}.
\subsection{Parallel Partitioning with MapReduce}
Our parallelization approach is based on following two insights.
First, spatial partition algorithms involve some kind of sorting based on a
derived spatial value, and MapReduce can perform such task very efficiently.
We can tweak the shuffle-and-sort phase of MapReduce to perform such task for
(almost) free.
Second, as different regions of a spatial dataset can be partitioned
independently, rather than changing the algorithms for parallelization,
we can run the partition algorithms on different regions of
the dataset in parallel. Although the generated partition layout may be
different from the one generated by a single thread partitioning program, it
is acceptable as long as the partitioning is reasonably well.

\SetStartEndCondition{ }{}{}%
\SetKwProg{Fn}{function}{\string:}{}
\SetKwFunction{In}{in}
\SetKw{KwTo}{in}
\SetKwFunction{mapfun}{Map}%
\SetKwFunction{redfun}{Reduce}%

\begin{algorithm}
\KwIn{a set of spatial objects $R$ }
\KwIn{partition payload $b$ }

$S$ = sample\_for\_partitioning(R)\;

\Fn{\mapfun {$k$,$v$}}{
$anchor$  = getAnchor($v$)\;
$key$  = calculateKey($anchor$,$S$)\;
emit($key$ , $v$)\;
}
\BlankLine
\tcc{shuffle and sort by MapReduce}
\BlankLine

\Fn{\redfun {$k$,$v$}}{
\tcc{partition the bucket with algorithm X}
$P$ = genPartitionX($v$)\;
emit($P$)\;
}
\caption{MapReduce based spatial partition}
\label{algo:mapredpart}
\end{algorithm}

We propose following approach for MapReduce-based parallelization of spatial
partition algorithms. First, similar to Hadoop Terasort
\cite{terasort}, we sample the dataset to generate an anchor point list which
will be utilized in the partition function of MapReduce for partition
assignment. In the Map phase we calculate a spatial ordering anchor,
such as geometrical center or Hilbert Curve value, and generate a key based on
the sample points generated previously. Next, the MapReduce framework will
partition the objects into groups based on their anchor location and sorts
them on the anchor value. At this point, dataset is roughly partitioned into
large spatial partitions. Later in section \ref{sec:experiments}, we will
discuss issues related to this coarse level partitioning.
In the reduce phase, each reducer will work on a single large partition, and
further partitions them into smaller partitions. Algorithm
\ref{algo:mapredpart} gives a sketch of this approach.

\subsection{Partitioning on Sampled Data}
Efficiency of a partition algorithm is subject to a number of factors such as algorithm runtime
complexity, dataset characteristics and size.
In relational database systems, sampling is used in various tasks to avoid
full dataset processing. For example, typical histogram construction
algorithms work on a small fraction of sampled data, thus avoiding the
expensive full dataset statistics. Such approach is shown to be practical and
efficient for query processing and dataset approximation.
Therefore it is natural to ask that if we can generate a spatial partition schema on a sampled dataset,
which reasonably approximates a full dataset partitioning.

Specifically, given a sampling ratio $\gamma$, we uniformly sample the dataset to get a reduced dataset of size $\gamma|R|$,
and run a partition algorithm on the this reduced dataset. Then, we map the generated partition layout onto the original dataset for final partition assignment and boundary object replication. Sampling ratio is the main control variable in the sampling based approaches. If the sampling ratio is too low, the resulting partition
quality may suffer. On the other hand,
if the sampling ratio is unnecessarily high, the partition efficiency may
suffer while the partition quality is only marginally improved. We explore those issues
later in Section \ref{subsec:expsampling}.

One problem with sampling based
partitioning is that some approaches fail to generate an effective spatial
partitioning on sampled dataset. For example, HC and STR generates the
partition regions that may not cover the entire spatial universe (Fig. \ref{fig:exhc} and \ref{fig:exstr}) and the partition region MBRs
are tight. In such case, the resulting partitions from
the sampled dataset can not be used without further fix. How to adapt those
approaches for spatial partitioning on sampled dataset is a problem we are planning to explore in our future work.

\section{Experimental Evaluation} \label{sec:experiments}
\subsection{Experimental Setup}
We use Amazon EMR for our benchmarking tasks. For single thread benchmarking
of partition algorithms, we use a large memory physical machine that comes
with 128 GB memory. For spatial join query scalability tests, we
use general purpose extra-large instance as our core and task nodes. Each
extra-large instance is equipped with $15$ GB memory, $4$ virtual cores and
$4$ disks with $1680$ GB storage ($4\times420$ GB). The Amazon Machine Images (AMI) version we used
for the cluster nodes is 3.0.2. Amazon S3 is used as the primary data storage for data serving.

\subsection{Datasets and Queries}
\noindent\textbf{OpenStreetMap (OSM) Dataset.} OpenStreetMap\cite{osm} is a collaborative mapping project that aims to create a free editable map of the world. It contains spatial representation of geometric features such as lakes,
forests, buildings and roads. Spatial objects are represented by a specific data type such as points,
lines, and polygons. We downloaded the dataset from the official website, and parsed it into WKT format for Hadoop-GIS to process.
The table schema is fairly simple, and it has roughly $70$ columns. We use the polygonal representation table
which contains more than $87$ million spatial objects.

\noindent \textbf{Imaging (PI) Dataset.} This dataset comes from an in-silico analysis of pathology images,
by segmenting boundaries of micro-anatomic objects such as nuclei and tumor regions, represented as polygons. Boundaries of spatial objects
are validated, normalized, and represented with WKT format. We have a set of 18 images (44GB) from a
brain tumor study at Emory University Hospital. Each of these images contains $0.5$ million spatial objects on average.

\noindent\textbf{Benchmark Query.} We use a \emph{spatial join} query to empirically evaluate the impact of a specific spatial partitioning  on the query performance.
Spatial join query is a very expensive query type that used in many benchmarking tests to evaluate system performance. Processing of a spatial join query requires evaluation of the spatial
predicate against all possible record pairs, and therefore is very time consuming.
The spatial join predicate we used in the join query test is \verb|st_intersects|, and the \verb|SQL| expression of
this query was illustrated in \cite{aji13hadoopgis}.

\subsection{Parameters and Metrics}
\noindent\textbf{Partition Payload.} The two datasets, OSM and PI, are from different application domains,
and they have different characteristics. Therefore, using the same parameter to partition both datasets may be problematic. For example, if we partition the smaller dataset with an expected payload of $c$ -- a perfect parameter for this dataset that yields best query performance, it might be a too fine granular partitioning for the larger dataset.
To be able to make the results comparable, we define the partition payload relative to the dataset size. We use a wide range of \emph{fractions} that will be multiplied with the dataset size to obtain the actual partition payload. Table \ref{table:param} shows those numbers.
\begin{table}[htb]
\centering
{\small
\begin{tabular}{|c|c|c|c|c|c|c|c|c|c|c|c|c|c|}
\hline
$f$  & 0.001 & 0.005 & 0.01 & 0.02 & 0.05 & 0.1 & 0.2  & 0.5  & 1.0  &  5.0  \\ 
\hline
\end{tabular}
\caption{Partition Parameter: Fraction ($\times 10^{-2}$) }\label{table:param}
}
\end{table}

\noindent\textbf{Boundary Object Ratio.}
We define a simple metric to study the relationship between partition granularity and partition
quality in terms of boundary objects. For a dataset $R$ that partitioned into $k$ partitions
$P=\{p_{1},p_{2},.. p_{i} .. p_{k}\}$, we define the boundary object ratio as:
\begin{equation}
  \lambda= \frac{\sum^{k}_{i=1}|p_{i}|}{|R|} - 1
\end{equation}
$\lambda$ is a real value that lies in the interval $\left[0,\infty\right)$.
If a spatial partitioning does not induce any boundary objects,
the value of $\lambda$ would be $0$.

\subsection{Comparison of Partition Quality}\label{subsec:stat}
Before we evaluate the partition results with real queries, we present some
statistical properties of the generated partitions which can provide us
insights on the partition algorithm behavior and quality.
\subsubsection{Partition Balance}
\begin{figure}[htb]
  \centering
  \includegraphics[trim = 2mm 1mm 2mm 1mm,clip]{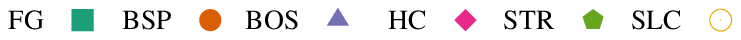}
  \subfigure[osm]{\includegraphics[width=0.49\linewidth]{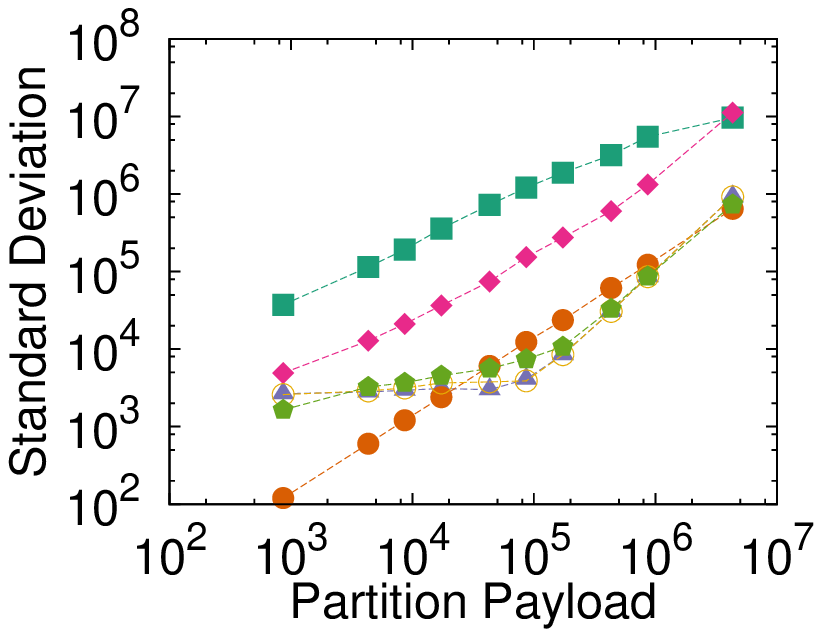}}
  \subfigure[pi]{\includegraphics[width=0.49\linewidth]{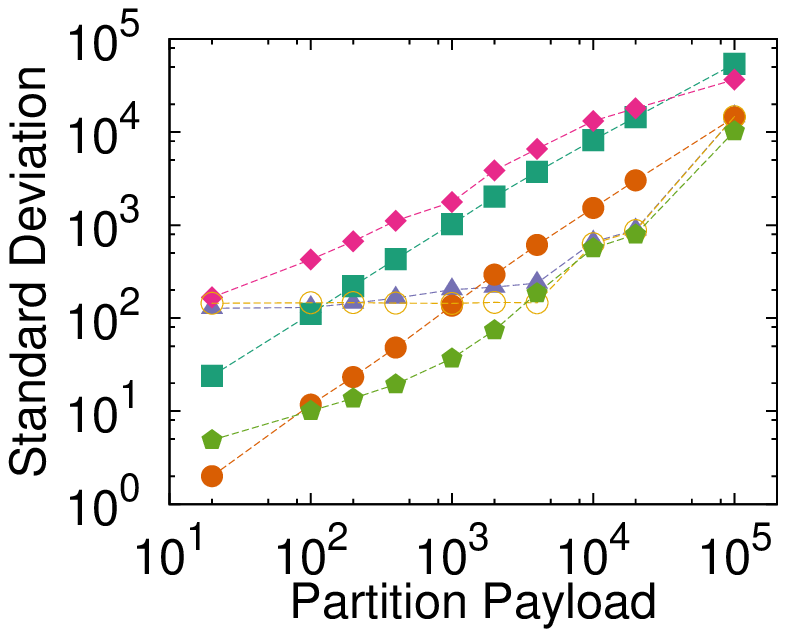}}
  \caption{Standard deviation of partition results}
  \label{fig:statstddev}
\end{figure}

Fig. \ref{fig:statstddev} shows standard deviation of generated partitions for different
partition algorithms on two datasets. Here, we use standard deviation as a measure of partition skewness.
The horizontal axis represents the expected partition payload -- a granularity value that we use to partition the
datasets. The vertical axis represents the standard deviation of generated
partition payloads.
Two conclusions can be made from the figure. First, as the partition
granularity increases, the skew tends to increase very quickly for all
methods. Therefore, a very coarse level spatial partitioning should be
avoided for parallel processing tasks that suffer from data skew.
Second, not surprisingly, FG generates significantly skewed partitions
compared to other approaches.

If we compare the same approach across two datasets with the same parameter setting,
we will find that the partitions generated from the OSM dataset is more skewed than the partitions generated from the PI dataset.
That means that, overall, the inherent skew in OSM is much severe than the PI dataset.
Furthermore, the FG partitioning for PI dataset is considerably better 
than the FG partitioning of OSM dataset. Therefore, we can conclude that, 
for a evenly distributed dataset,
the FG approach can generate a reasonably well partitioning. However, if the dataset is highly skewed,
FG approach may generate a very low quality partitioning.

Adaptive approaches, such as STR, BOS and SLC, should be able to handle certain level of
data skew as they can make smarter data oriented partition decision. We can see from the figures that corresponding lines
for those approaches are relatively flat until the partition granularity gets large. However, as the partitions get larger, the adaptability
of those algorithms also approaches their limitations.

One interesting result we did not expect to see is that partitions generated by HC approach are also as skewed as FG partitions, and for the PI dataset HC is not even as good as FG.
As HC approach is a data oriented approach that traditionally used for bulk loading spatial indexes, it is surprising that the partitioning from HC has such high imbalance.

\begin{figure}[htb]
  \centering
  \includegraphics[trim = 2mm 1mm 2mm 1mm,clip]{stats/statlegend.eps}
  \subfigure[osm]{\includegraphics[width=0.49\linewidth]{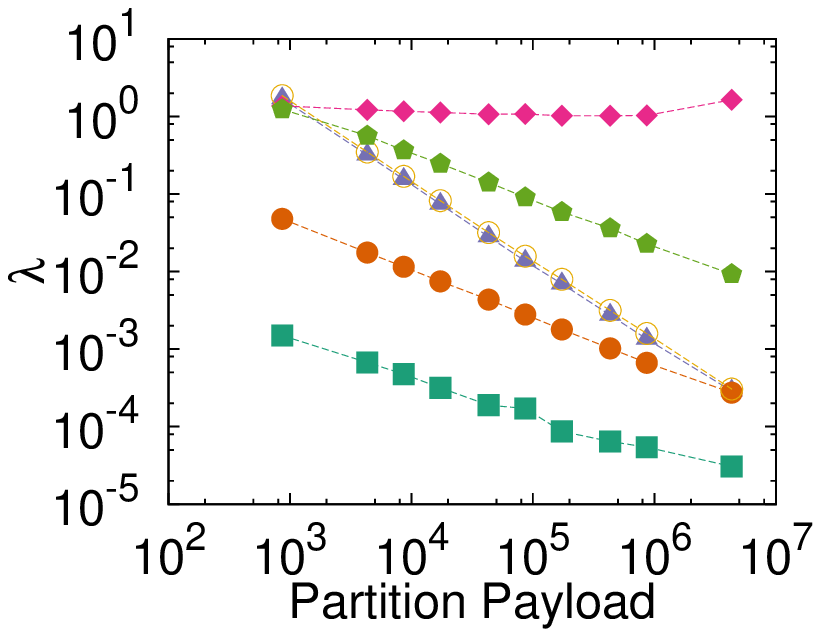}}
  \subfigure[pi]{\includegraphics[width=0.49\linewidth]{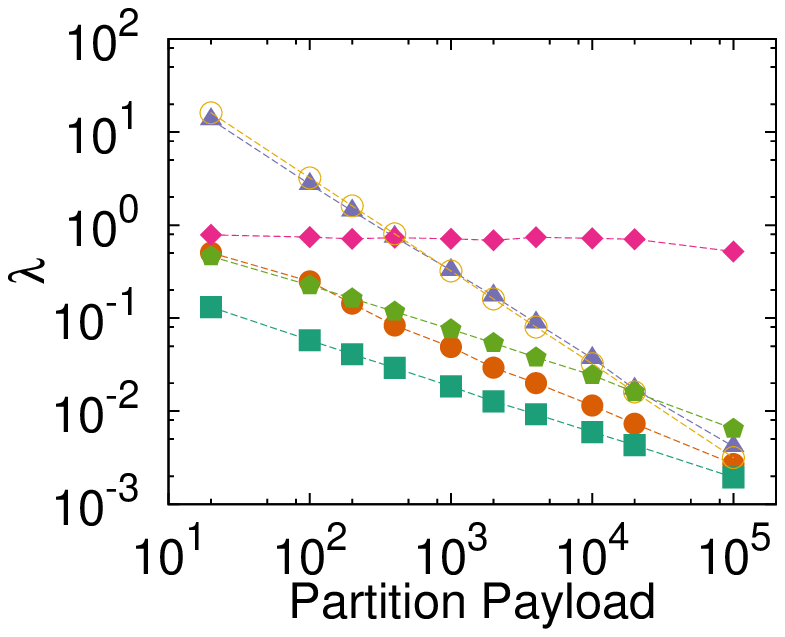}}
  \caption{Ratio of boundary objects}
  \label{fig:statratio}
\end{figure}

\subsubsection{Boundary Objects}
Fig. \ref{fig:statratio} shows the ratio of boundary objects generated by different algorithms.
We can see the overall trend that, for both datasets, as the partition granularity increases the ratio of boundary objects decreases.
FG seems to be a good algorithm if our main objective is to have less boundary objects. However, as both figures show, a very fine granular
partitioning is problematic as it significantly increases the dataset size, and in certain cases such increase can be dramatic.
For example, if we look at the $\lambda$ value for the first horizontal axis data point in
Fig. \ref{fig:statratio} (a), for Strip partitioning (SLC) the boundary object ratio is $1.86$, whereas the same data point value
is $16.1$ in Fig. \ref{fig:statratio} (b). Such a large increase in data size is certainly not acceptable, and we can conclude that a very
fine granular partitioning is not a practical approach for large scale query processing.

Interestingly, in both figures, the lines for the \emph{slicing} approaches, SLC and BOS,
have higher slopes than other approaches. This indicates that, for those partitioning algorithms,
even a slight increase in the partition payload can contribute to significantly less number of boundary objects.
Therefore, in practice, those partition methods should be configured to generate a relatively larger size partitions
so that the number of boundary objects are reasonably small.

\subsection{Effects of Partitioning on Query Performance}
In this section, we empirically evaluate partition algorithms on different configurations
to study how a specific partitioning affects the query performance, and investigate the
relationship between partition granularity and query performance.
The experiments are performed on a $50$ node Amazon AWS MapReduce cluster,
and general purpose AWS instances are used as compute nodes and storage nodes.
Each experiment is conducted three times, and average of those three runs is used to account for performance variations in cloud environment.

\begin{figure}[htb]
  \centering
  \includegraphics[trim = 2mm 1mm 2mm 1mm,clip]{stats/statlegend.eps}
  \subfigure[osm]{\includegraphics[width=0.49\linewidth]{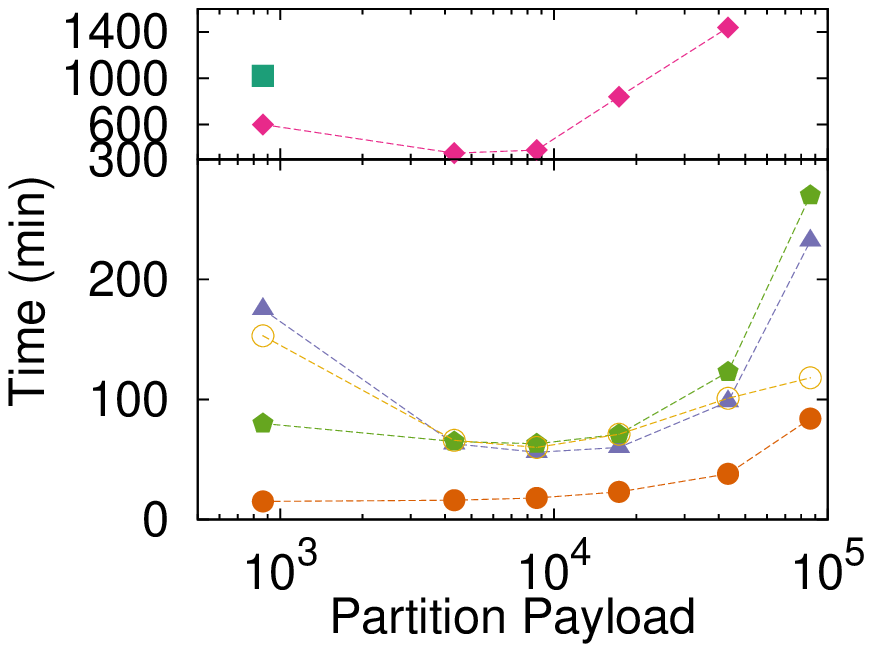}}
  \subfigure[pi]{\includegraphics[width=0.49\linewidth]{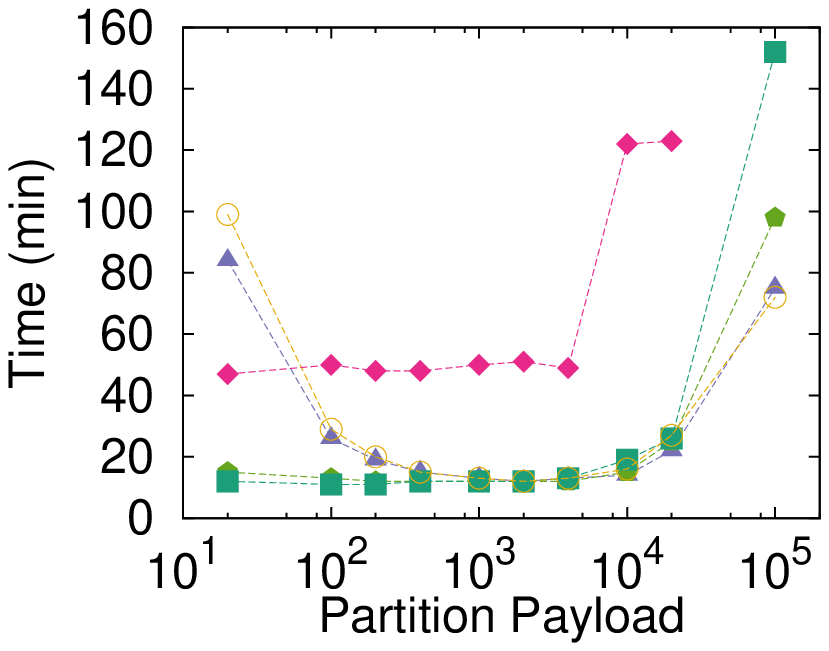}}

  \caption{Spatial join query performance}
  \label{fig:perfjoin}
\end{figure}

Fig. \ref{fig:perfjoin} shows the performance of the spatial join query on two datasets.
The horizontal axis represents the partition granularity,
and vertical axis represents the query performance.
Clearly, neither a very fine or very coarse partitioning yields the optimal query performance.
For a fine granular partitioning, the main cause can be attributed to
the high boundary object ratio which not only increases the I/O overhead, but also the extra computation overhead.
For a very coarse granular partitioning, however, the root cause is the data skew between partitions.


Recall that, in section \ref{sec:background}, our cost analysis framework suggests that there is a point of
optimal partition granularity that yields best query performance. The performance numbers on both datasets support such case.
As the figures show, overall, query performance is close to the optimal in mid-range of horizontal axis,
and performance starts to degrade as the partition granularity increases.
However, if we compare different algorithms over a wide range of partition granularities,
it is difficult to generalize such statement. Specifically, BSP and STR have relatively better
performance on a wider range of partition granularities, and the performance starts to suffer only after the partition granularity becomes too large.
This can be attributed to the properties of these algorithms that they can adaptively handle data skew and boundary objects.

\noindent\textbf{Performance variance between datasets.} In Fig. \ref{fig:perfjoin} (a), the performance of different approaches are tiered.
FG and HC have similar performance, and their performance are almost orders of magnitude worse than other approaches
(due to the long query runtime, we only report one data point for FG).
While performance of HC is still the worst on PI dataset as shown in Fig. \ref{fig:perfjoin} (b),
performance of FG, however, is almost optimal for most cases.
Clearly, specific characteristics of a dataset are contributing to such difference.
Our observation indicates that PI dataset consists of large number of \emph{small} objects
that are fairly evenly distributed across space, whereas OSM dataset consists of \emph{variety of objects}
of all sizes that are clustered around a number of hotspots.
If we simply consult to the statistical propoerties from the previous subsection \ref{subsec:stat},
we can also see that FG partitioning of PI
dataset is less skewed compared to the OSM dataset. Moreover, the number of boundary objects from
FG partitioning is very small on all partition granularity. Due to those reasons, on PI dataset, FG partitioning
achieves a balanced partitioning for ``free'', and has an unfair advantage over other approaches.

\subsection{Partition Efficiency}
In this subsection we study the partition efficiency of different algorithms.
To perform a fair comparison, the time for reading the dataset from the disk, and writing the partition results to the
disk is not included in the performance measurement. The performance time only includes the time for deriving
the actual partition boundaries after the dataset is loaded into the main memory of a single machine.

\begin{figure}[htb]
  \centering
  \subfigure[osm]{\includegraphics[width=0.49\linewidth]{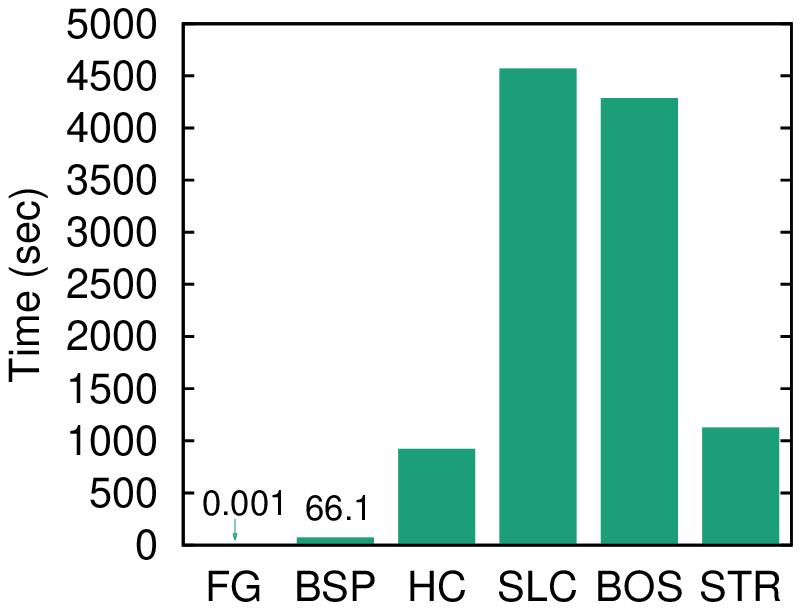}}
  \subfigure[pi]{\includegraphics[width=0.49\linewidth]{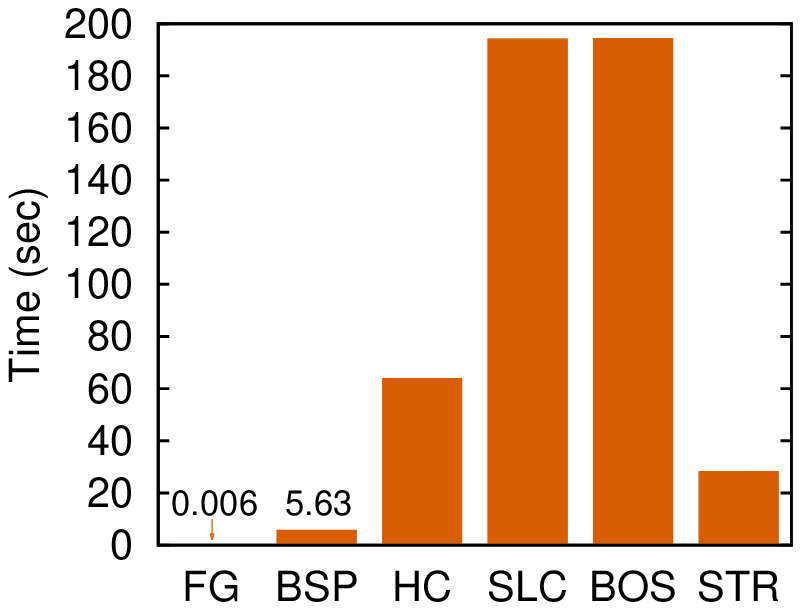}}
  \caption{Spatial partition performance}\invpiclable
  \label{fig:partperf}
\end{figure}

Fig. \ref{fig:partperf} shows the runtime cost of partition algorithms on two datasets.
Depending on the actual runtime performance, algorithms can be roughly categorized into three
categories -- fast (FG, BSP), medium (HC, STR), and slow (SLC, BOS). For both datasets,
FG partition has the lowest runtime cost which is only in the range of milliseconds, and BSP has
the second best performance. However, other four algorithms require considerable amounts of time to generate partitions.
Specifically, the space slicing approaches -- SLC and BOS, require more than an hour to derive a partitioning on OSM.
This is mainly due to the nature of the algorithms that SLC and BOS not only sort the dataset on one dimension,
they also perform lots of boundary object examination. The main cost of HC is the Hilbert Curve calculation
and sorting based on the curve value. The performance of the algorithms on different datasets is
roughly similar, with the exception of HC that has a slightly slower performance on the PI dataset compared to the OSM.

\begin{figure}[htb]
  \centering
  \includegraphics[trim = 2mm 1mm 2mm 1mm,clip]{stats/statlegend.eps}
  \subfigure[osm]{\includegraphics[width=0.49\linewidth]{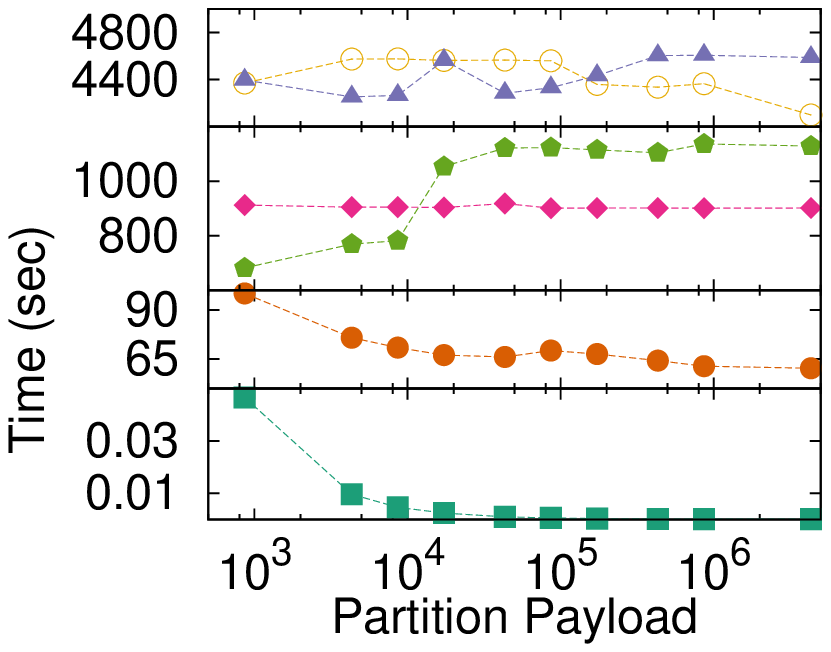}}
  \subfigure[pi]{\includegraphics[width=0.49\linewidth]{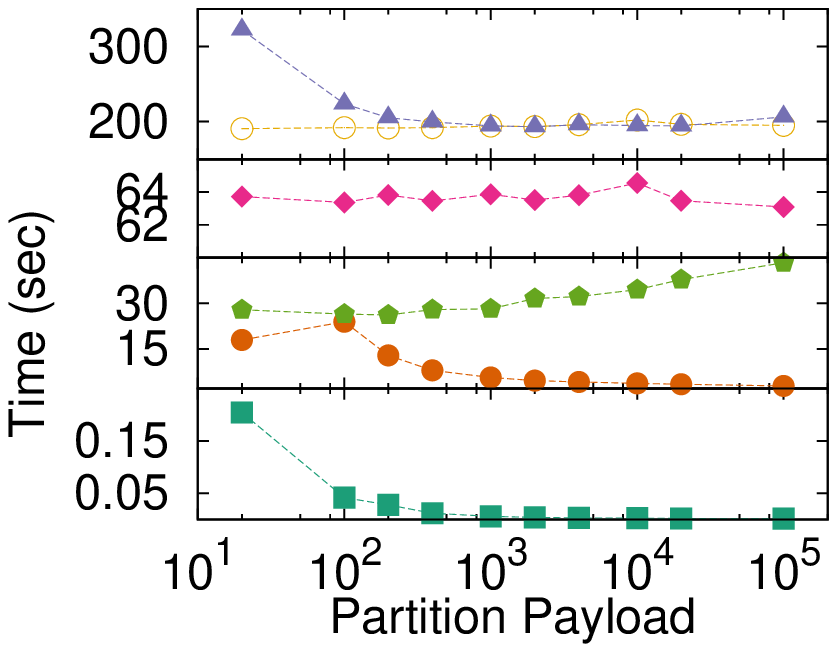}}
  \caption{Spatial partition performance variance}
  \label{fig:partperfvar}
\end{figure}

Figure \ref{fig:partperfvar} shows the runtime performance of the algorithms over different partition granularity.
While the performance of the algorithms do not depend too much on the partition granularity, there are noticiable
differences. Intuitively, a finer granularity partitioning entails more cpu cycles, and therefore it is expected
that algorithms run slower for small payload values. Performance numbers of FG and BSP show 
such tendency.
However, depending on the algorithm and dataset characteristics such hypothesis may not hold true.
For example, the main cost in HC partitioning comes from calculating and sorting the spatial objects
based on the Hilbert curve value. Regardless of partition granularity, such cost is constant.
Therefore, as the figure shows, performance of HC does change with partition granularity.
Interestingly, STR has lightly degraded performance on a larger partition granularity on OSM dataset.
The specific reasons are not completely clear to us, and we are planning to investigate such problem in future work.

If we compare relative performance of the algorithms across the two datasets, the lines for PI dataset
is more smooth and predictable. For example, on OSM dataset, SLC and BOS have an irregular runtime
performance over different partition payloads. However, those algorithms do not exhibit the same
behavior in the PI dataset. Given the dataset characteristics we discussed earlier,
we can conclude that dataset characteristics have implications for the algorithm performance.

\subsection{Parallel Partitioning with MapReduce}\label{subsec:expmapred}
Spatial partitioning is a time consuming process, and, as the performance numbers in previous subsection show, it may take hours.
This has motivated us to develop MapReduce based spatial partitioning for improved query performance and spatial ETL process.
To test efficiency and scalability of our MapReduce based parallel partitioning approach,
we modified and tested selected set of four partitioning algorithms, namely BSP, SLC, BOS and STR.
The rationale in such selection is that, (1) parallelization of FG and HC is straightforward,
and (b) they generate suboptimal partitioning in most cases. Here, we select a set of 
three expensive spatial partitioning approaches (SLC, BOS, STR) to experiment. While BSP is 
reasonably fast, we also include it in our experiments to compare its performance with 
other approaches.  
Experiments are also performed on the Amazon EMR,
and unlike the performance measurement in previous subsection, here the runtime performance
includes both I/O cost and computation cost.

\begin{figure}[htb]
  \centering
  \subfigure[scalability]{\includegraphics[width=0.49\linewidth]{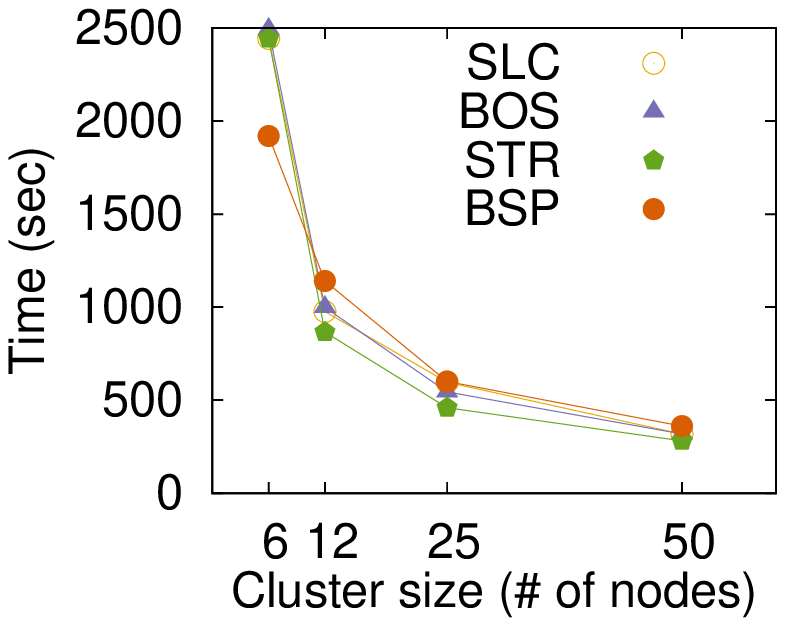}}
  \subfigure[performance]{\includegraphics[width=0.49\linewidth]{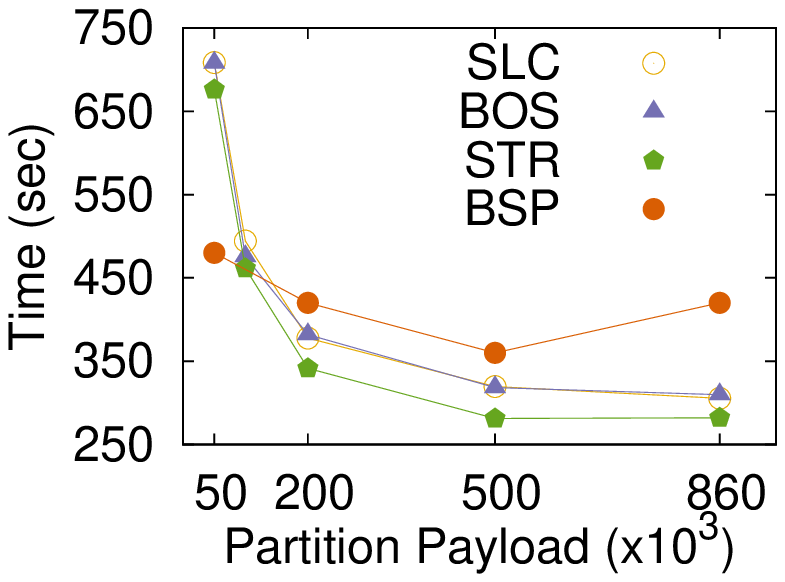}}
  \caption{Parallel partitioning performance}
  \label{fig:mapredpart}
\end{figure}

Fig. \ref{fig:mapredpart} (a) shows a scalability chart for the three MapReduce based parallel partitioning approaches on OSM dataset.
The horizontal axis represents the number of nodes used for parallelization, and the vertical axis represents the partition runtime.
The performance is measured with a top level coarse partition granularity of $500000$.
While this number seems to be arbitrary, our experiments show that the scalability is not affected by the coarse partitioning granularity.
As the figure shows, the MapReduce based partitioning approach is very scalable and efficient.
With the increased cluster capacity, the runtime performance improves almost linearly.
With parallelization, the partition efficiency of the algorithms increased by an order of magnitude. For example, the runtime
of BOS decreased from $4000$ seconds to merely $300$ seconds. Although the algorithms have very different runtime
performance on a single thread implementation, the performance after parallelization seems to be homogeneous.

Recall that our parallelization algorithm performs partitioning in two steps.
The top level coarse partitioning for parallel partitioning,
and bottom level partitioning in which the coarse partitions are re-partitioned with specific spatial partition algorithms.
Each step involves a partitioning granularity parameter which controls partition size.
To study the effects of those parameters on parallel partitioning performance, we perform two seperate experiments.
In the first experiment, we fix the coarse top level partitioning granularity and test the runtime performance with different bottom level partitioning granularity. Not surprisingly, the performance difference between different parameters are too little to be significant, and consequently we can conclude that the bottom level partitioning granularity has no noticeable effect on parallel partitioning performance.

In the second experiment, we fix the bottom level partitioning,
and change the top level partition granularity. Figure \ref{fig:mapredpart} (b),
shows performance variations of parallel partitioning for different partition
granularity. We can see that as the top level partitioning granularity gets
coarser, the performance gets better.
Our profiling of the parallel algorithms provides folowing explanation.
Like Terasort \cite{terasort}, the parallelization algorithms use a sampled data file
for assigning the spatial objects into separate partition groups which has a
global total ordering. In a finer granularity top level spatial partitioning,
the total order based partition group assignment becomes the major bottleneck.
Interestingly, the visualization of the partition boundaries show that spatial partition results from a larger top level partitioning has more resemblance to the partition results from a single threaded approach.

\subsection{Spatial Partitioning with Sampling}\label{subsec:expsampling}
\begin{figure}[hbt]
  \centering
  \subfigure[BSP-sttdev]{\includegraphics[width=0.49\linewidth]{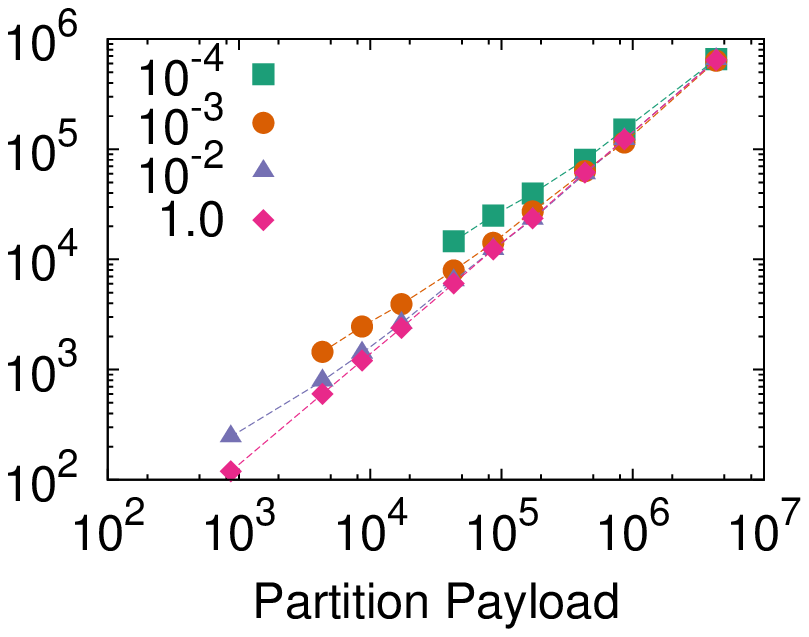}\label{fig:z}}
  \subfigure[BSP-ratio]{\includegraphics[width=0.49\linewidth]{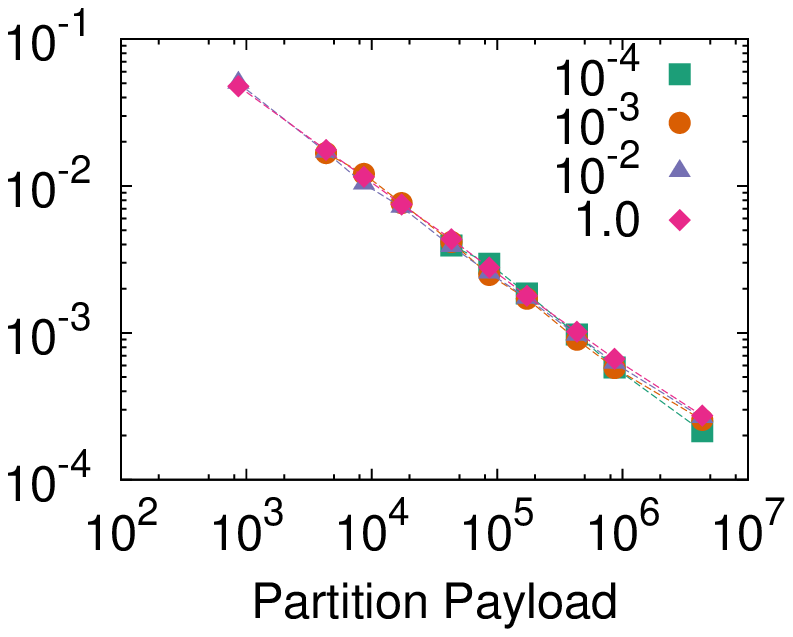}\label{fig:zz}}
  \subfigure[SLC-stddev]{\includegraphics[width=0.49\linewidth]{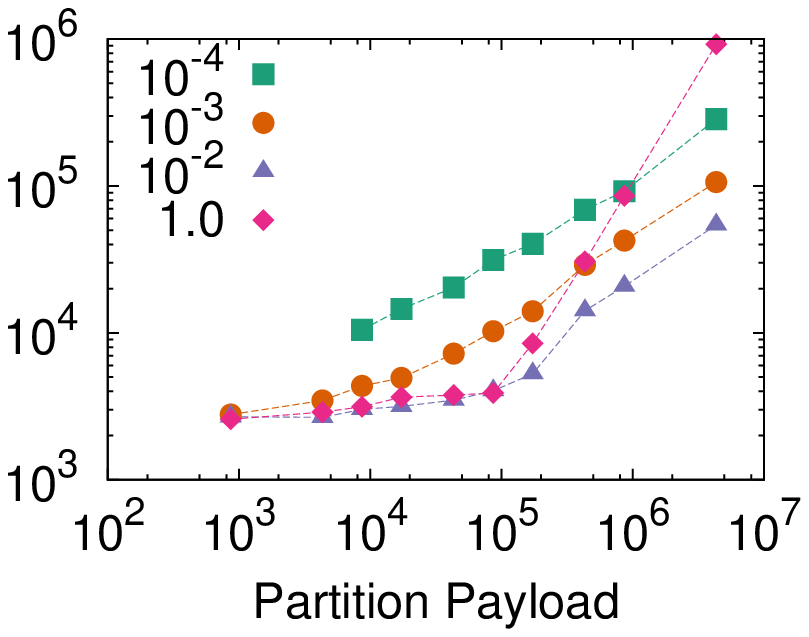}\label{fig:zzz}}
  \subfigure[SLC-ratio]{\includegraphics[width=0.49\linewidth]{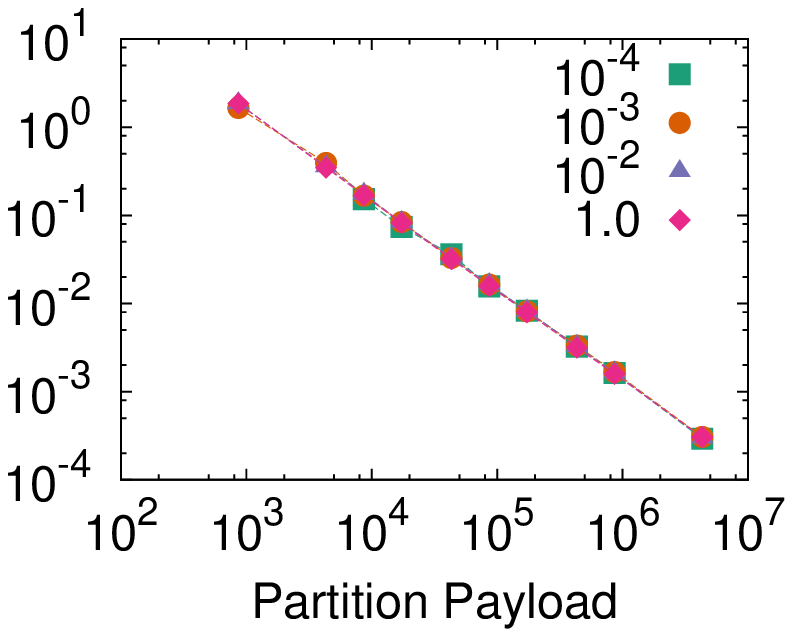}\label{fig:zzzz}}
  \subfigure[BOS-stddev]{\includegraphics[width=0.49\linewidth]{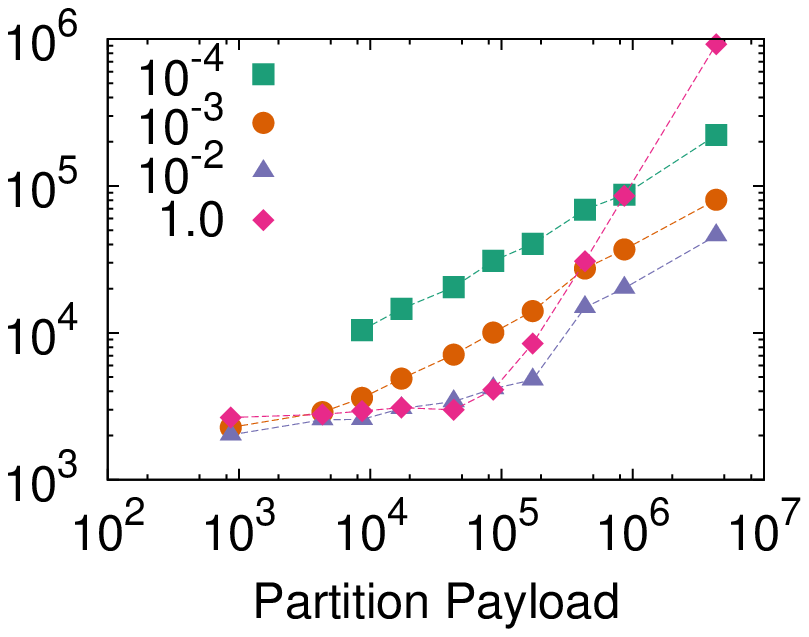}\label{fig:zzzzz}}
  \subfigure[BOS-ratio]{\includegraphics[width=0.49\linewidth]{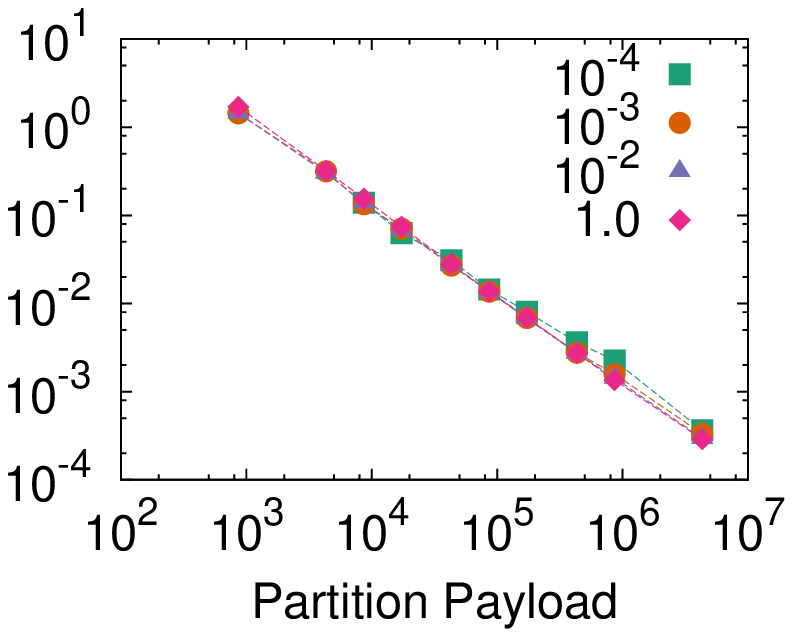}\label{fig:zzzzzz}}
  \caption{Quality of partitions generated by sampling based approaches}
  \label{fig:samplingpartitionquality}
\end{figure}

Fig. \ref{fig:samplingpartitionquality} shows a statistical evaluation of
three sampling based partitioning approaches on the OSM dataset.
The figures on the left column show the standard deviation -- measure of skewness --
of generated partitions,
and the figures on the right column show boundary object ratio.
The full dataset is sampled with different sampling rate
(shown in the legend of the figures), the resulting partitions from the sampled
dataset are compared against the the partitioning generated from the full dataset.
The sampling rate of $1.0$ represents full dataset partitioning.
From the figures we can see that sampling can be a very effective approach for
spatial
partitioning.

Intuitively, higher the sampling rate, the better we can preserve data distribution,
and consequently the partitioning on the sampled dataset is of higher quality.
If we look at the figures on the left column, we can see that partitions
generated with higher sampling rate are less skewed compared to lower sampling
rate partitioning. However, depending on the algorithm, partition skew can be different.
For example, as BSP implicitly try to avoid a skewed partitioning,
the impact of higher sampling rate is not significant.
Whereas in SLC and BOS, higher sampling rate seems to be always beneficial.
There is a minor exception to this case. Specifically, in SLC and BOS,
if the partition payload is reasonably large, sampling based approaches can generate
a less skewed partitioning than the full dataset partitioning.
This is particularly interesting, and it has important implications for certain application
scenarios. First, by using a sampling based approach we can significantly reduce the partition time.
Second, aside from the improved performance, we can actually obtain a less skewed partitioning
with the minor limitation of large partition size.
Interestingly, the ratio of boundary objects generated by sampling based partition approaches is not
completely dependent on the sampling ratio. Overall, the sampling based partitioning approaches generate more boundary objects compared to the full dataset partitioning, although the variation is not significant. 

\section{Related Work}\label{sec:relatedwork}
To the best of of our knowledge, this is the first work that studies the spatial data 
partition problem in detail. Data partition problem is discussed extensively in the 
context of database systems in the last few decades \cite{ceri1982horizontal,scheuermann1998data}. Fixed grid partition and its variations are used for spatial join processing in \cite{patel97building,zhou1998data}. Le et al. \cite{le2013optimal} studied the problem of finding optimal splitters for large 
interval data. More recently, MapReduce based systems emerged as an effective solution to Spatial Big Data challenges \cite{evans2014spatial}. 
HadoopGIS \cite{aji13hadoopgis} is a spatial data warehousing system that is based on a general spatial query processing framework. The system uses SQL as the query language, and integrated into Hive\cite{thusoo09hive}. SpatialHadoop \cite{eldawy2013demonstration} is an extension of 
Hadoop for spatial query processing, and it extends Pig \cite{eldawypigeon} at the query language layer. Ray et al. \cite{ray2013parallel} proposed a spatial data analysis infrastructure that uses a combination of cloud environment and relational database systems. Authors also briefly discussed a hybrid 
approach that uses Hilbert Curve and space partitioning for spatial join processing.

Spatial histogram construction is extensively studied in database settings, and it is widely used for approximate query processing. The main goal of spatial histogram construction is to partition the multi-dimensional data into buckets (most often a bucket represents a rectangular region), where data within buckets is uniformly distributed. In that sense, spatial histogram generation is relevant to spatial partitioning, but not the same. In \cite{muthukrishnan1999rectangular}, authors have showed that computing the non-overlapping rectangular partitioning with near-uniform data distribution within buckets is NP-hard. One of the pioneering works is \cite{muralikrishna1988equi}, in which authors proposed to extend the concept of equi-depth histogram to multidimensional data. An in-memory data structure \emph{hTree} is designed for storing the histograms. It constructs non-overlapping partitioning of multidimensional space based on object frequencies. However location of objects are not considered for histogram construction, which may result in skewed histograms. \emph{MinSkew} histogram \cite{acharya1999selectivity} is proposed to remedy some of the disadvantages of hTree. 
Specifically, the authors proposed two construction strategies. The first approach has two phases. In the first phase, the algorithm tiles the spatial universe  into uniform regular grids and stores the number of intersecting spatial objects for each tile. Then based on the tiling, a recursive binary space partitioning (BSP) is used for histogram construction. Authors experimentally observed that a fixed-size tiling is sensitive to the size of the queries (high grid resolution favors small sized queries and vice versa), and proposed another approach \emph{MinSkew-Progressive-Refinement} which can utilize multi-resolution tiling.

Spatial histogram construction is extensively studied in spatial database settings, and it is widely used for approximate query processing. The main goal of spatial histogram construction is to partition the multi-dimensional data into buckets (most often a bucket represents a rectangular region), where data within buckets is uniformly distributed. GenHist \cite{gunopulos2000approximating} is a recent approach which can identify high density regions for real valued attributes. However, in GenHist bucket rectangles may overlap, and the buckets can be contained in other buckets. It uses a fixed-size grid as the basis of histogram construction. More recently, an approach called STHist \cite{roh2010hierarchically} is proposed to generate density aware histograms. In the basic STHist algorithm, decision about whether the region is dense is made by applying a sliding window over all dimensions, by approximating the frequency distribution by a marginal distribution. 
In the advanced variant called STForest, the algorithm first computes coarse partitions according to the object skew, and then applies a sliding window algorithm to them. 
STHist has a time complexity of $\mathcal{O}(n^2)$ for 2-dimensional and $\mathcal{O}(n^3)$ for 3-dimensional data.

A convenient approach to obtain a spatial histogram is to generate it using a spatial index structure like R-Tree \cite{rtree1984guttman}, R${^{*}}$-Tree \cite{BeckmannKSS90}, R$+-$Tree \cite{sellis1987rplus} etc. \emph{RK-Hist} \cite{eavis2007rk} is an example of such approach which is based on R-tree bulk-loading procedure. The data is presorted according the Hilbert space-filling-curve. After the leaf nodes are generated, a histogram can be generated by packing nodes according to the sorting order in equi-sized histogram buckets. However, this may not necessarily generate a good partitioning. Specifically, for approximately uniformly distributed data equi-sized partitioning wastes buckets for regions with a high object density and produces high overlap between buckets. Therefore, the authors proposed a greedy algorithm utilizing a sliding window of pages along the Hilbert order. The algorithm is parametrized with a number of buckets that should be considered for a split. A bucket-split is applied if it leads to an improvement according to the proposed cost function.  More recently, a new approach \emph{R-V} \cite{achakeev2012class} is proposed to overcome skewed-data distribution problem.

\inv
\section{Conclusion }\label{sec:conclusion}
A proper spatial partitioning schema is essential for optimal query performance and system efficiency for scalable
distributed spatial query processing.
In this paper, we formally introduce the spatial partition problem, and present a comprehensive study of
six different partitioning algorithms. We categorize the algorithms along three dimensions, and provide
a systematic evaluation of the algorithms on two real world datasets from different domains. We also propose
parallelization methods to improve the efficiency of spatial data partition process, and explore sampling based
partitioning as an alternative for fast spatial partitioning. Our study reveals several insights on how
partitioning effects query performance and what factors should be considered for effective spatial partitioning.
The results provide practical guidelines for designing spatial partitioning for large scale parallel spatial
query processing.

\section{Acknowledgements }\label{sec:acknowledgements}
This work is supported in part by NSF IIS 1350885, by NSF ACI 1443054,  by NLM R01LM009239 and NCI \\
1U24CA180924-01A1. We are also grateful for the support from Amazon AWS in Education, Pitney Bowes, and Google Summer of Code.

\balance

\small
\bibliographystyle{abbrv}
\bibliography{hadoopgis}

\end{document}